\documentclass[]{AO4ELT}  

\usepackage{microtype}
\usepackage[sorting=none]{biblatex}
\usepackage{amsmath,amsfonts,amssymb}
\usepackage{graphicx}
\usepackage{pst-all} 
\usepackage[colorlinks=true, allcolors=blue]{hyperref}
\makeatletter         
\def\@maketitle{
\includegraphics[width = 170mm]{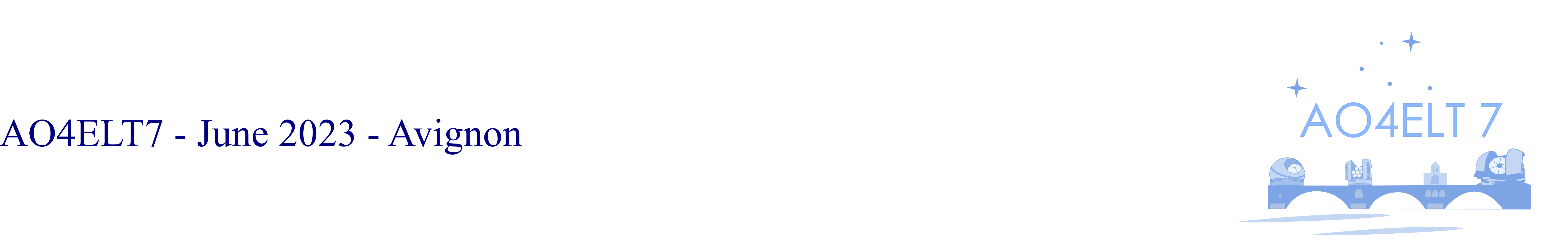}\\[8ex]
\begin{center}
{\Huge \bfseries \sffamily \@title }\\[4ex] 
{\Large  \@author}\\[4ex] 
\@date
\end{center}}

\title{CiaoCiao WFS: sensing phase discontinuities at the Extremely Large Telescope}

\author[a]{Giulia Carlà}
\author[a]{Lorenzo Busoni}
\author[a]{Simone Esposito}
\author[a]{Guido Agapito}
\author[b]{Ronald Holzl{\"o}hner}
\affil[a]{INAF, Osservatorio Astrofisico di Arcetri, Largo Enrico Fermi 5, I-50125 Firenze, Italy}
\affil[b]{European Southern Observatory (ESO), Karl-Schwarzschild-Str. 2, 85748 Garching, Germany}

\authorinfo{Further author information: (Send correspondence to G.C.)\\G.C.: E-mail: giulia.carla@inaf.it}

\begin{document} 
\maketitle

\begin{abstract}
The upcoming extremely large telescopes will have to deal with the so-called “pupil fragmentation” effect: for what concerns the Extremely Large Telescope (ELT), the presence of thick spider legs supporting the secondary mirror may induce unseen phase discontinuities across the pupil sectors that could limit the performance of the adaptive optics correction. In this context, we propose a wavefront sensor (WFS), the \textit{CiaoCiao WFS}, consisting in a rotational shearing interferometer to sense phase differences between the pupil sectors. In this work, we present the \textit{CiaoCiao WFS} concept and the first analyses carried out through numerical simulations. In particular, we analyze the performance of such a wavefront sensor in case the phase discontinuities are induced by low-wind effect during observations with the Multiconjugate adaptive Optics Relay For ELT Observations (MORFEO).
\end{abstract}


\section{INTRODUCTION}
\label{sec:intro}
The upcoming 25-39-m class of extremely large telescopes (ELTs) is currently under design and realization with the aim to provide astrophysical observations in near-infrared wavelengths with unprecedented resolutions. Due to their large sizes, the pupils of such telescopes are inevitably segmented and partially shaded by thick support structures. This leads to the effect known as ``pupil fragmentation'' that can severely degrade the performance of the adaptive optics (AO) systems that will equip the ELTs to reach the diffraction limit. For what concerns the 39-m Extremely Large Telescope (ELT) \cite{Tamai20}, such an effect is generated by the presence of 50-cm-wide spider legs supporting the secondary mirror that creates thick shadows in the pupil, thus a fragmentation of the pupil in six disjoint segments whose separation is larger than the typical Fried parameter. Due to this configuration, the wavefront sensor (WFS) subapertures under the spiders are blind and thus the reconstructed wavefront may present discontinuities between the pupil sectors. 
If not sensed and controlled, phase discontinuities can severely degrade the AO performance even during the best atmospheric turbulence conditions. 

Low Wind Effect (LWE) can be one of the contributors to the discontinuities between pupil sectors that has first been observed in 2015 at the Very Large Telescope during the commissioning of SPHERE \cite{Sauvage15,Sauvage16}. The sharp steps that were observed in the optical path difference (OPD) around the VLT spiders have then been explained through the cooling of the spiders due to the radiative transfer between the spiders and the sky background  \cite{Holzlohner21}: the air flowing near the spider cools down and, in case of low-wind conditions ($<$ 5 m/s), is not well mixed; this effect leads to temperature gradients that induce discontinuities in the OPD across the spiders. Several strategies have been proposed for a passive mitigation of LWE-induced phase discontinuities \cite{Milli18}, but no solutions foreseeing direct measurements have been defined so far.

The Multiconjugate adaptive Optics Relay For ELT Observations (MORFEO) \cite{Busoni23}, that will deliver multiconjugate adaptive optics (MCAO) correction to the first-light Multi-AO Imaging Camera for Deep Observations (MICADO) \cite{Davies21} of the ELT, will be equipped with Shack Hartmann (SH) WFSs. Being slope sensors, SH WFSs are not able to measure phase jumps across the spiders of the ELT since the distance between the sectors is larger than the atmospheric coherence length. Pupil fragmentation can then be a major issue for MORFEO performance.

In this context, we propose the \textit{CiaoCiao WFS} concept \cite{Busoni22}, evolved from the work presented in Ref.~\cite{Holzlohner19}, to sense phase discontinuities across the pupil sectors of the ELT: by taking advantage of the rotational symmetry of the ELT pupil, we aim at using a rotational shearing interferometer \cite{Malacara} to make interference between two pupil sectors separated by a spider in order to sense phase differences between adjacent sectors. Though an extended analysis of the \textit{CiaoCiao WFS} is being carried out - from both a numerical and an experimental point of view - and is going to be provided in a future publication, in this work we want to focus on the first results of the numerical analysis concerning the capability of the \textit{CiaoCiao WFS} to measure LWE-induced phase differences between the ELT pupil sectors.

In Section \ref{sec:ciao_ciao_intro}, we present the \textit{CiaoCiao WFS} concept; in Section \ref{sec:software}, we go through the steps of the Python software that we setup to perform numerical simulations in order to analyze the performance of the \textit{CiaoCiao WFS}; in Section \ref{sec:results}, we show first results on \textit{CiaoCiao WFS} performance, including the measurements of phase differences induced by LWE on MORFEO residual wavefront.

\section{The CiaoCiao WFS}
\label{sec:ciao_ciao_intro}

\begin{figure}[htbp]
    \centering
    \includegraphics[width=0.7\linewidth]{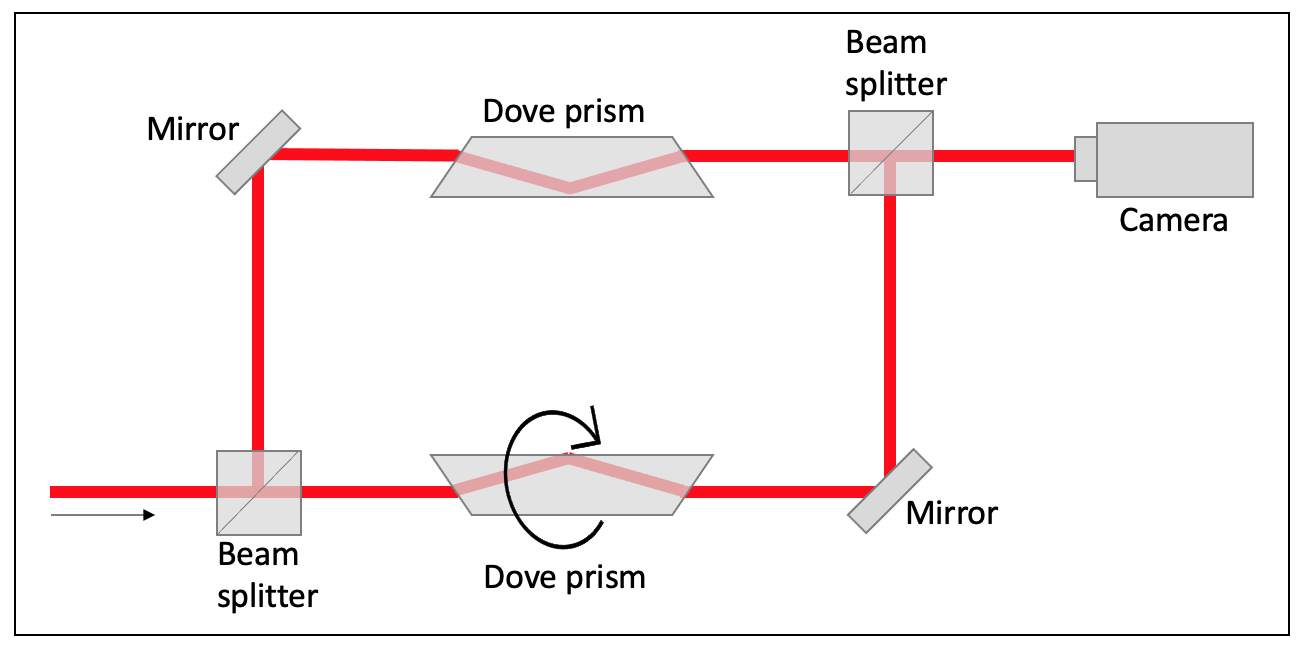}
    \caption{Sketch of a simplifed setup of the \textit{CiaoCiao WFS}. A cube beam splitter divides the incident wavefront into two beams and sends them into two interferometer arms. Each arm includes a Dove prism: the first one (bottom) rotates the wavefront with respect to the wavefront in the other interferometer arm; the second one (top) is stationary and is used to compensate the changes in the optical path. The rotated beam and the reference one overlap again at a second beam splitter and the interferograms are acquired on a camera.}
    \label{fig:basic_setup}
\end{figure}

We propose to use a rotational shearing interferometer as WFS - the \textit{CiaoCiao WFS} - to measure phase differences across the spiders of the ELT pupil.
The basic principle is to have a setup (e.g. a configuration based on a Mach-Zender interferometer as schematized in Fig.\ref{fig:basic_setup}) that simultaneously acquires the image of the pupil and the same image rotated by an angle $\Delta\theta$, and that makes interference between the two.

\begin{figure}[htbp]
    \centering
    \includegraphics[width=\linewidth]{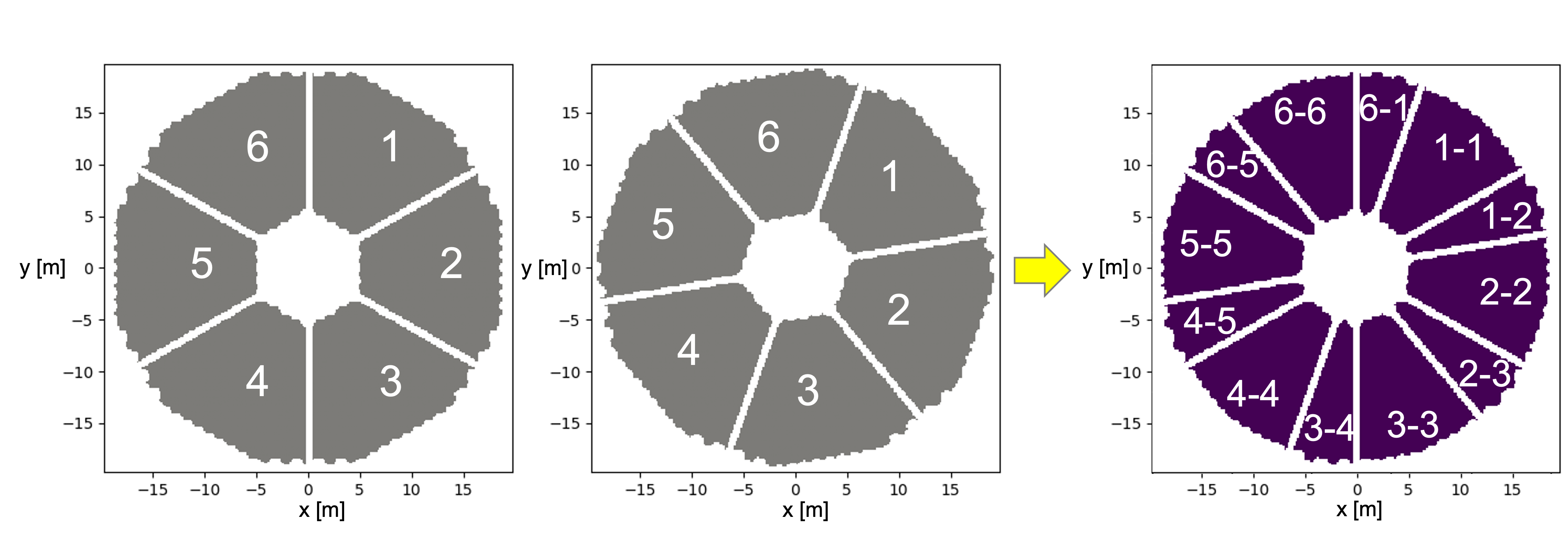}
    \caption{Left: image of the pupil; center: image of the pupil rotated by 20$^{\circ}$; right: overlapping sectors after interference.}
    \label{fig:basic_concept}
\end{figure}

Considering the electric fields of the two beams:
\begin{equation}
\begin{split}
    \label{eq:electric_fields}
    E_1(\textbf{r}, t) &= A_1(\textbf{r}) \cos(\omega t + \varphi_1(\textbf{r})) \\ 
    E_2(\textbf{r}, t) &= A_2(\textbf{r}) \cos(\omega t + \varphi_2(\textbf{r})) \, ,
\end{split}
\end{equation}
the measured intensity is the time-averaged squared magnitude of the interfered electric field:
\begin{equation}
\begin{split}
    \label{eq:intensity}
    I(\textbf{r}) &= \langle |E_1(\textbf{r}, t) + E_2(\textbf{r}, t)|^2 \rangle_T \\
    &= \dfrac{I_1(\textbf{r})}{2} + \dfrac{I_2(\textbf{r})}{2} + \sqrt{I_1(\textbf{r}) I_2(\textbf{r})}\cos(\Delta\varphi(\textbf{r})) \, ,
\end{split}
\end{equation}
where
\begin{equation}
\begin{split}
    \Delta\varphi(\textbf{r}) &= \varphi_1(\textbf{r}) - \varphi_2(\textbf{r}) \\
    &= \varphi_1(r, \theta) - \varphi_1(r, \theta - \Delta\theta) \, .
\end{split}
\end{equation}
Thus, by inverting Eq.~\eqref{eq:intensity}, we can estimate the phase difference between two pupil points having the same radial coordinates and the azimuthal coordinates differing by the rotational angle $\Delta\theta$ that is defined from the setup of the shearing interferometer. If we average the measurements of $\Delta\varphi$ on the overlapping sectors, we get the estimate of the phase difference between them (Fig.~\ref{fig:basic_concept}):
\begin{equation}
    \label{eq:ave_phase_diff}
    \overline{\Delta\varphi} = \dfrac{1}{S} \int_0^R dr 
    \int_0^{\Delta\theta} d\theta \left[\varphi(r, \theta) - \varphi(r, \theta - \Delta\theta) \right] \, ,
\end{equation}
where $S$ is the overlapped area and $R$ is the pupil radius. Considering a sensing wavelength $\lambda$, it follows the measurement of the optical path difference (OPD) as 
\begin{equation}
    \label{eq:opd}
    \mathrm{OPD} = \dfrac{\lambda}{2\pi} \overline{\Delta\varphi} \, .
\end{equation}

It is worth noting that if $\Delta\theta$ is small, the \textit{CiaoCiao WFS} measures phase differences across the spiders; on the other hand, if $\Delta\theta$ is 60$^{\circ}$, adjacent sectors are completely overlapped in the interferogram and then the \textit{CiaoCiao WFS} measures the differential average phase between the ELT pupil sectors.

\section{Software implementation}
\label{sec:software}
We have written a Python software to perform numerical simulations to test the \textit{CiaoCiao WFS} performance. The software is based on POPPY \cite{poppy} that we use to create an optical model as stack of pupil layers. In our model we can include the ELT pupil aperture, the petaled M4, the turbulence, low order Zernike and also load maps (e.g. AO residual turbulence, LWE) from \textit{.fits} files.

The steps of a single simulation are:
\begin{itemize}
    \item setup two identical optical models and rotate one of them by $\Delta\theta$;
    \item generate interferograms with proper phase shifts for phase reconstruction - the software implements the four step phase shifting interferometry algorithm;
    \item extract phase difference maps and convert them to OPD maps;
    \item estimate the jumps across adjacent regions that have been overlapped - i.e. average OPD on the overlapping sectors;
    \item derive petal signals $\textbf{p}_m$ from the cumulative sum of the measured jumps $\textbf{j}_m$:
        \begin{equation}
        \begin{split}
            \label{eq:petals_offset}
            \textbf{p}_m &= - \left[j_1, j_1 + j_2, j_1 + j_2 + j_3, \cdots \right] \\
            &= - \left[\left(p_6 - p_1\right), \left(p_6 - p_1\right) + \left(p_1 - p_2\right), \left(p_6 - p_1\right) + \left(p_1 - p_2\right) + \left(p_2 - p_3\right), \cdots \right] \\
            &= \left[p_1 - p_6, p_2 - p_6, p_3 - p_6, p_4 - p_6, p_5 - p_6, 0 \right] \, ,
        \end{split}
        \end{equation}
        that is, in the algorithm the petals are estimated with an offset given by the signal on the 6-\textit{th} sector and the correction applied based on these measurements will introduce a global piston given by $p_6$. Of course, this is not an issue for what concerns the AO performance.
\end{itemize}

\section{Numerical simulations}
\label{sec:results}
We show first tests concerning the \textit{CiaoCiao WFS} measurement of pure pistons injected on the ELT pupil sectors; then, we focus on the effects produced by the LWE on MORFEO residual wavefront and on the capability of the \textit{CiaoCiao WFS} to compensate for such effects.

\subsection{Injection of pure pistons}
\begin{figure}[ht]
    \centering
    \includegraphics[width=1.\linewidth]{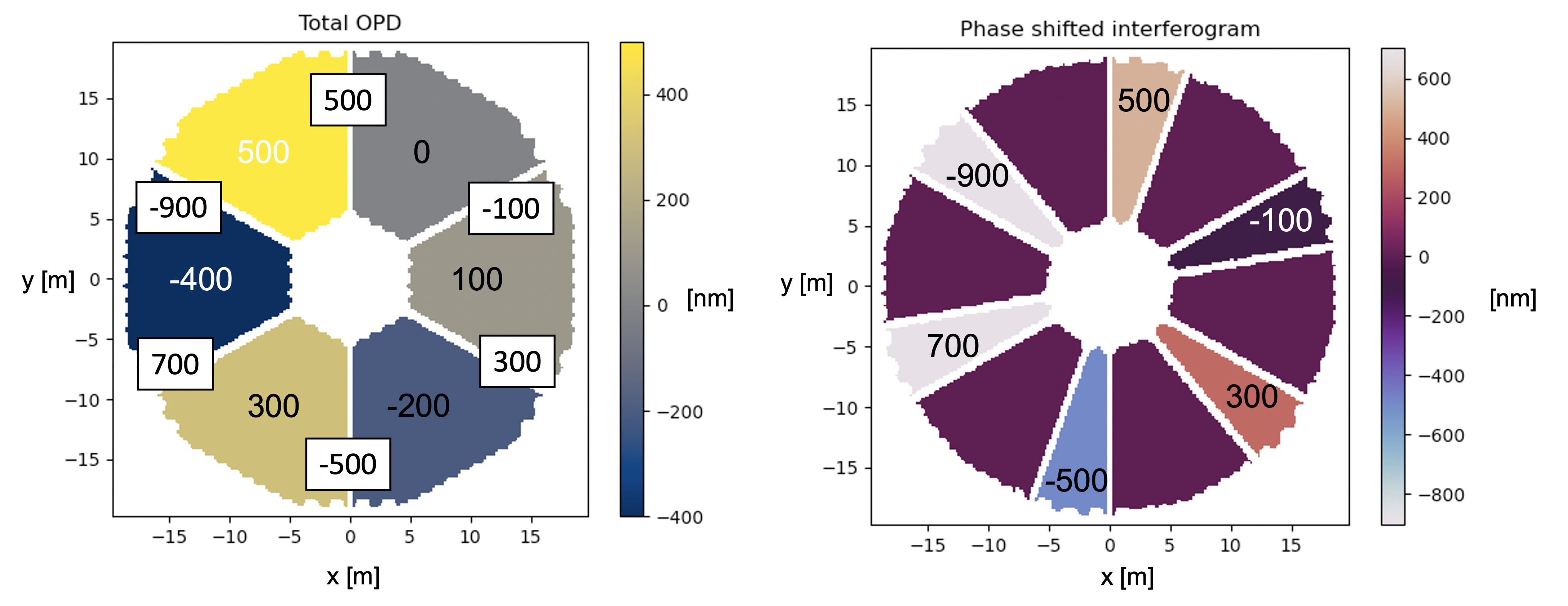}
    \caption{Left: OPD map showing the pistons injected on the ELT pupil sectors and the expected jumps (in the white boxes). Right: jumps measured by the \textit{CiaoCiao WFS} from the interferogram of the two pupil images having a differential rotation angle of 20$^{\circ}$.}
    \label{fig:test_pistons}
\end{figure}

We consider pure pistons applied on the ELT pupil sectors, as shown in Fig.~\ref{fig:test_pistons} (left plot). We assume the \textit{CiaoCiao WFS} sensing at $\lambda$=2.2 $\mu$m and with a rotation angle of 20$^{\circ}$. Considering the ordering of the sectors shown in Fig.~\ref{fig:basic_concept}, the injected pistons are:
\begin{equation}
    \textbf{p} = [0, 100, -200, 300, -400, 500] \: \mathrm{nm} \, .
\end{equation}
The \textit{CiaoCiao WFS} measures the following jumps across the sectors (right plot of Fig.~\ref{fig:test_pistons}):
\begin{equation}
    \textbf{j}_m = [500, -100, 300, -500, 700, -900] \: \mathrm{nm} \, ,
\end{equation}
that lead to the estimation of the following signals on the sectors (from Eq.~\eqref{eq:petals_offset}):
\begin{equation}
    \textbf{p}_m = [-500, -400, -700, -200, -900, -0] \: \mathrm{nm} \, .
\end{equation}
Thus, the measurements from the \textit{CiaoCiao WFS} are correct up to a global piston of -500 nm; indeed, as shown in Section \ref{sec:software}, they have an intrinsic offset given by the signal on the 6-\textit{th} sector. In this case, the correction from the \textit{CiaoCiao WFS} will leave a residual global piston, that clearly does not represent an issue for the AO performance.

\begin{figure}[ht]
    \centering
    \includegraphics[width=1.\linewidth]{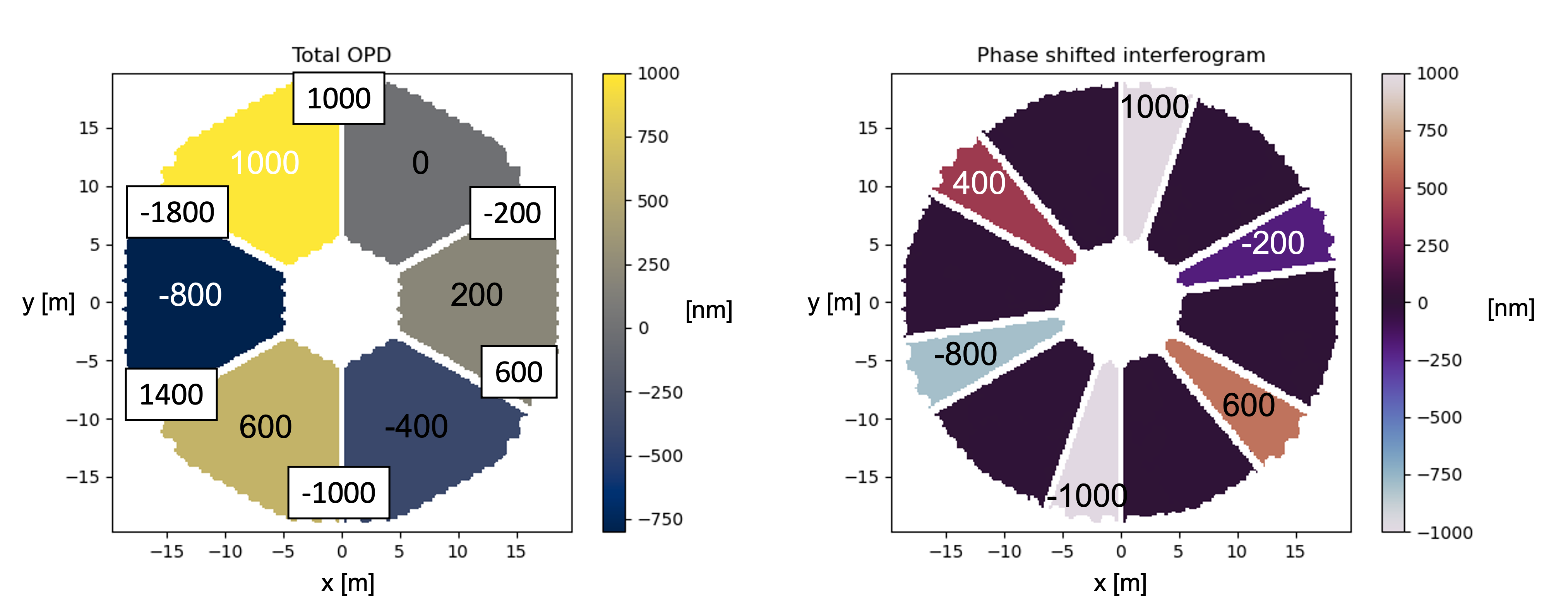}
    \caption{Left: OPD map showing the pistons injected on the ELT pupil sectors and the expected jumps (in the white boxes). Right: jumps measured by the \textit{CiaoCiao WFS} from the interferogram of the two pupil images having a differential rotation angle of 20$^{\circ}$. The injected pistons here lead to jumps exceeding the dynamic range [-$\lambda$/2; +$\lambda$/2].}
    \label{fig:test_pistons_large}
\end{figure}

In Fig.~\ref{fig:test_pistons_large}, we apply pistons up to 1000 nm on the ELT pupil sectors:
\begin{equation}
    \textbf{p} = [0, 200, -400, 600, -800, 1000] \: \mathrm{nm} \, .
\end{equation}
Though the pistons are within the dynamic range of the interferometric measurements, the jumps exceed such range and the \textit{CiaoCiao WFS} measures:
\begin{equation}
    \textbf{j}_m = [1000, -200, 600, -1000, -800, 400] \: \mathrm{nm} \, ,
\end{equation}
that is, the last two measurements are wrong by -/+$\lambda$. This leads to the following petals estimation:
\begin{equation}
    \textbf{p}_m = [0, 200, -400, 600, 1400, 1000] \: \mathrm{nm} \, ,
\end{equation}
where the 5-\textit{th} petal measurement is not correct.

This simple test shows the issue given by the modulo $\lambda$ ambiguity if carrying out the OPD measurements with monochromatic light. Considering a sensing wavelength of 2.2 $\mu$m - i.e. the longest wavelength that we can access in the framework of the ELT/MORFEO -, the dynamic range is limited to $\pm$ 1100 nm and jumps larger than this will lead to wrong measurements from the \textit{CiaoCiao WFS}. In this context, we are investigating the possibility to implement multi-wavelength reconstruction techniques.

\subsection{MORFEO residuals including the compensation for LWE}
In this section, we report first results of the analysis concerning the measurement of LWE-induced differential phase on MORFEO residuals. 

\begin{figure}[ht]
    \centering
    \includegraphics[width=0.6\linewidth]{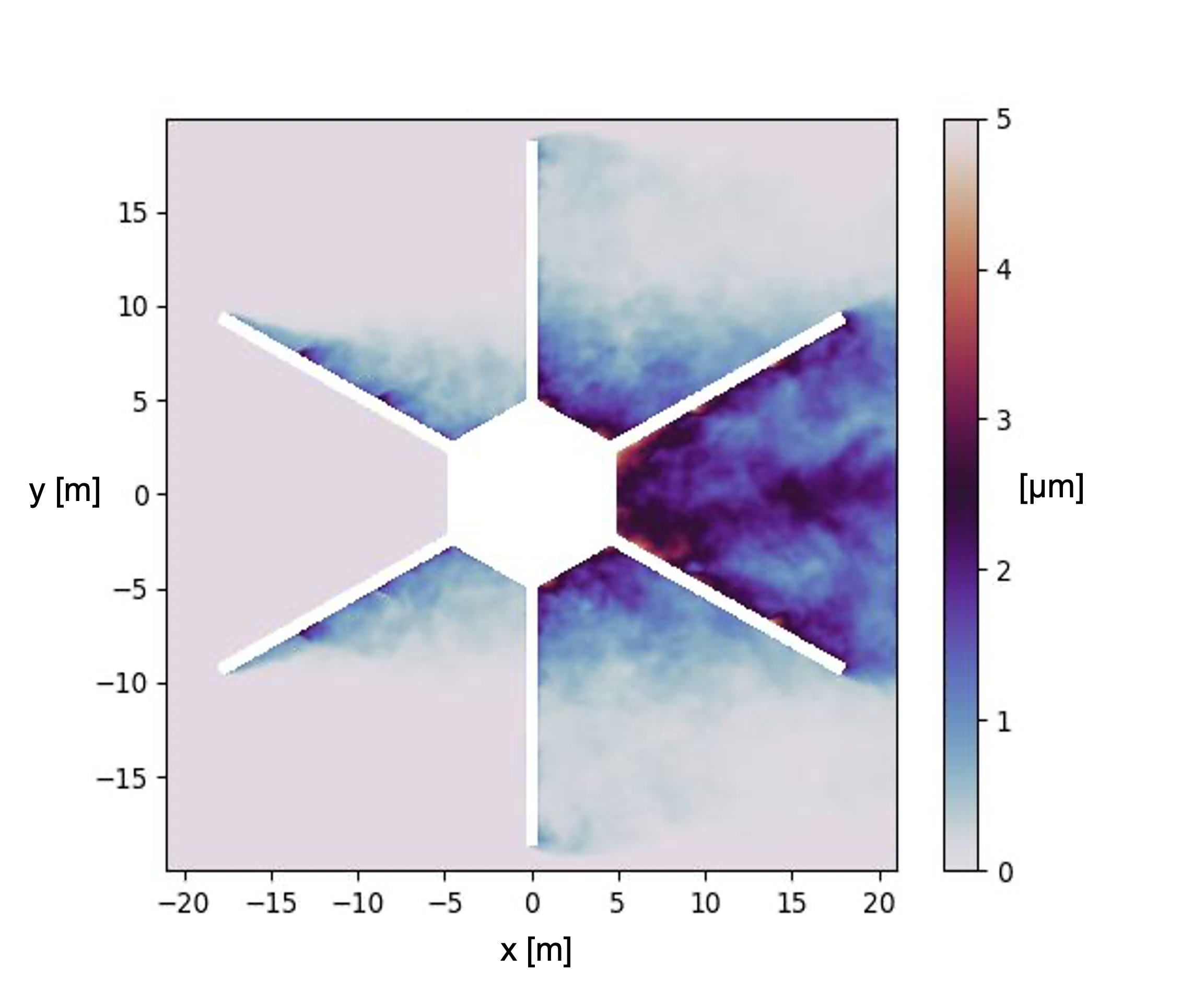}
    \caption{OPD map from CFD simulations of LWE \cite{Martins22} (one instantaneous frame). The wind is directed along the \textit{x} axis and its speed is 0.5 m/s.}
    \label{fig:lwe_map}
\end{figure}

\begin{figure}[ht]
    \centering
    \includegraphics[width=0.7\linewidth]{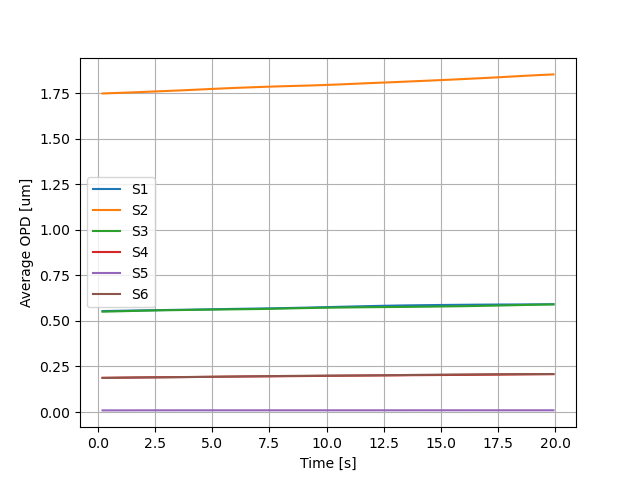}
    \caption{Average OPD on each ELT pupil sector as a function of time. Colors represent each sector and the ordering is the same as the one of Fig.~\ref{fig:basic_concept}. S1 and S3 are superimposed, as well as S4 and S6.}
    \label{fig:lwe_vs_time}
\end{figure}

In Fig.~\ref{fig:lwe_map}, we show the OPD maps from LWE that we used and that have been derived from the Computational Fluid Dynamics (CFD) simulations that are reported in Ref.~\cite{Martins22}. The simulations consist of 99 steps of 0.2 s each and the wind speed is 0.5 m/s. As a first analysis, we extract the average OPD produced by LWE on each sector and we show the variation over time in Fig.~\ref{fig:lwe_vs_time}: the average OPD on the ELT pupil sectors determined by LWE is characterized by a slow temporal evolution and the differential average phase between adjacent sectors is up to $\sim$1.2 $\mu$m.

\begin{figure}[htbp]
    \centering
    \includegraphics[width=\linewidth]{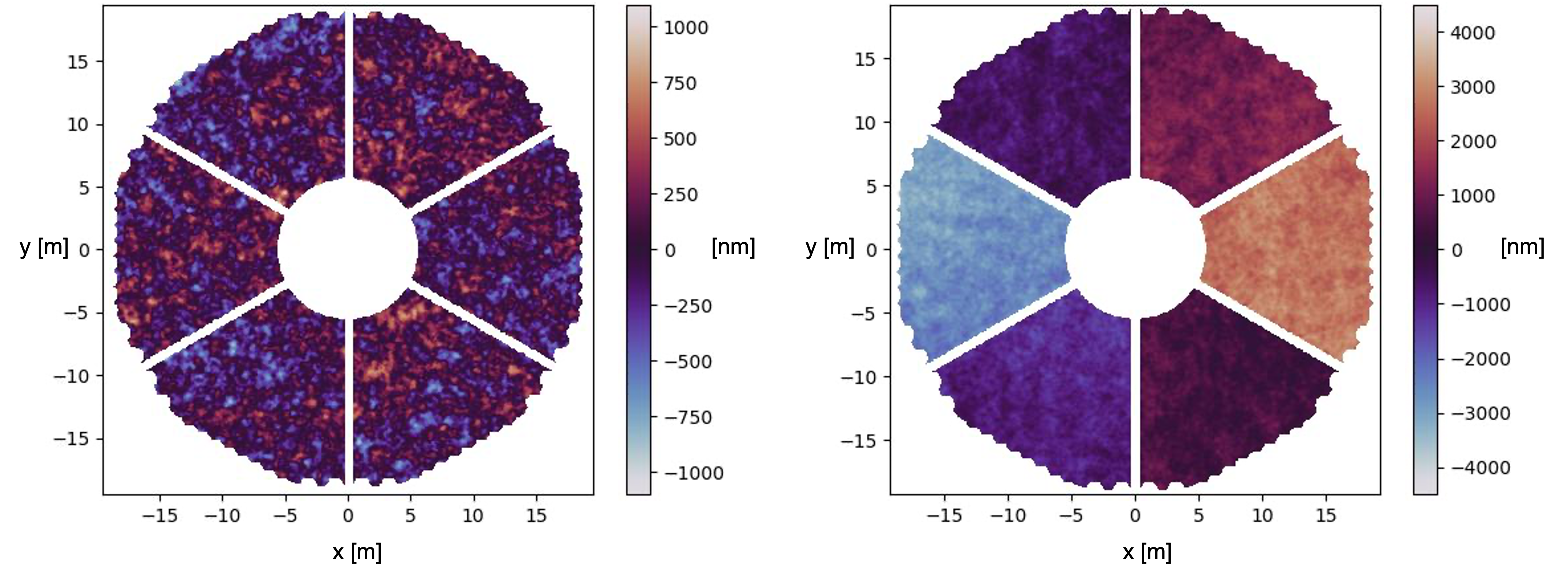}
    \caption{Left: MORFEO residual wavefront from PASSATA simulations (one frame) that include only atmospheric turbulence as input. Right: MORFEO residual wavefront obtained from PASSATA simulations (one frame) when also LWE is included.}
    \label{fig:morfeo_with_lwe}
\end{figure}

For what concerns the OPD maps of MORFEO residual turbulence, we got them from PASSATA end-to-end simulations \cite{passata}. The simulations consist of 2 s at 500 Hz and implement a typical MORFEO configuration - 3 NGSs on 55" equilateral asterism, 6 LGSs on 45" asterism, control and reconstruction based on POLC-MMSE algorithms - with atmospheric turbulence characterized by a seeing of 0.644 arcsec (i.e. median conditions). The OPD maps of MORFEO residual turbulence are shown in Fig.~\ref{fig:morfeo_with_lwe} (left image) and present a flat wavefront on which we measure an RMS of 255 nm.

If we include LWE as input together with atmospheric turbulence within PASSATA simulations, we get the OPD maps shown in the right image of Fig.~\ref{fig:morfeo_with_lwe}: MORFEO residual wavefront appears now to be ``petalled'' and the residual RMS goes to 1744 nm, showing that the LWE can severely affect the operation of MORFEO.

\begin{figure}[htbp]
    \centering
    \includegraphics[width=\linewidth]{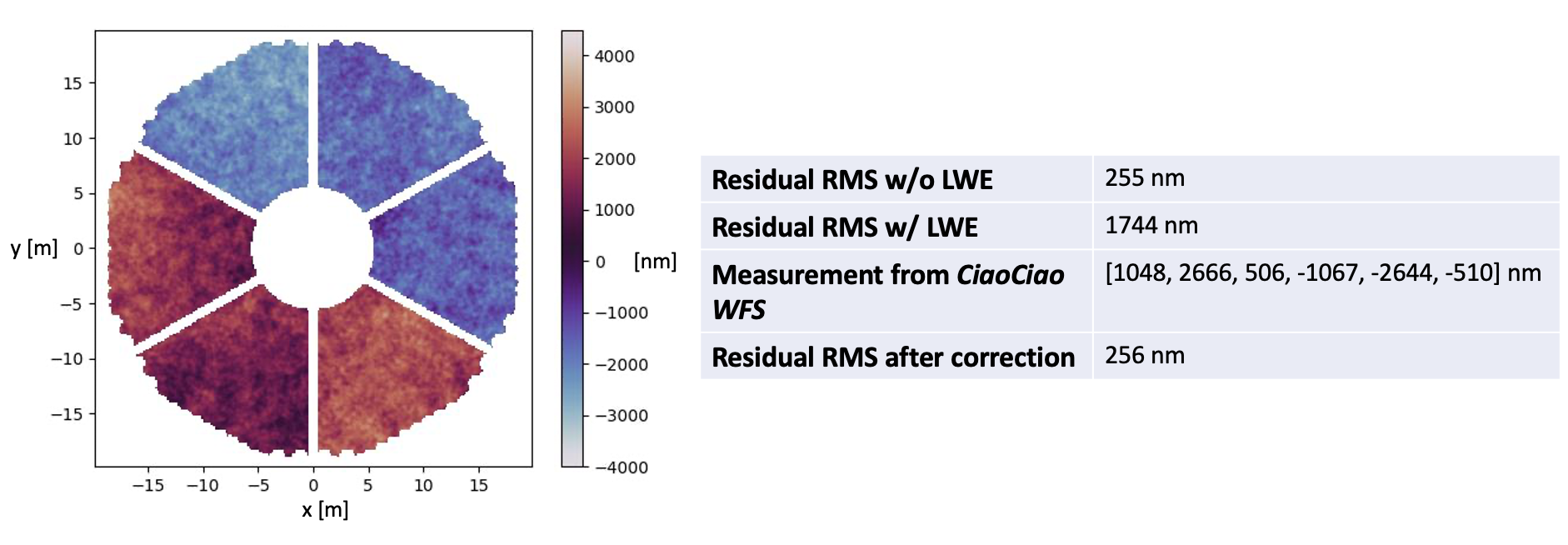}
    \caption{Left: Interferogram obtained with a rotation angle of 60$^{\circ}$ on MORFEO residuals including LWE. Right: Table reporting the RMS without and with LWE included in the simulations, the piston measurements from the \textit{CiaoCiao WFS} and the residual RMS after the correction is applied based on the \textit{CiaoCiao WFS} signals.}
    \label{fig:lwe_final_results}
\end{figure}

We test the \textit{CiaoCiao WFS} on MORFEO residual wavefront maps including LWE, in order to verify if it is able to restore the RMS expected for the considered MORFEO configuration: since the impact of LWE is to introduce differential pistons between the ELT pupil sectors on MORFEO residual wavefront, the effect could be indeed easily detected by the \textit{CiaoCiao WFS} having a setup with a rotation angle of 60$^{\circ}$. One issue is the dynamic range: from the left plot of Fig.~\ref{fig:morfeo_with_lwe}, we can measure jumps between adjacent sectors that exceed the range available with a 2.2 $\mu$m sensing wavelength. Therefore, we simulated a setup including a dual wavelength reconstruction \cite{Houairi09} in H band ($\lambda_1$ = 1.5 $\mu$m, $\lambda_2$ = 1.6 $\mu$m) that provides a synthetic wavelength of 24 $\mu$m. In Fig.~\ref{fig:lwe_final_results}, we show the interferogram (left) obtained with a rotation angle of 60$^{\circ}$ and the results of the simulation (right): the \textit{CiaoCiao WFS} measures pistons up to $\sim$ 2.7 $\mu$m that, used to apply a correction to MORFEO residual wavefront, bring the residual RMS back to 256 nm. 

Considering the above analysis, it is worth noting the following aspects that we aim to investigate:
\begin{itemize}
    \item the \textit{CiaoCiao WFS} should not have stringent requirements on sky coverage for what concerns the measurement of LWE-induced differential pistons. Indeed, as shown in Fig.~\ref{fig:lwe_vs_time} the temporal evolution of LWE is slow and long integration times could be used allowing for the sensing on faint stars;
    \item since LWE appears to "apply" six pistons on the ELT pupil sectors, we may not need a detector with high spatial resolution. We then intend to test different spatial samplings;
    \item a K-band WFS does not provide the dynamic range required by LWE-induced phase differences. We did first tests with dual-wavelength reconstruction and we want to investigate further multi-wavelength techniques.
\end{itemize}

\section{Conclusions}
We presented the \textit{CiaoCiao WFS}, a rotational shearing interferometer to sense phase discontinuities at the ELT. Thanks to the rotational symmetry of the ELT pupil, it could be a simple solution to tackle the issue of pupil fragmentation that, if not controlled, will severely affect the performance of the adaptive optics systems. Depending on the rotation angle of the shearing interferometer, the \textit{CiaoCiao WFS} can sense either phase discontinuities at the edge of the spiders or the differential average phase between the adjacent sectors. We focused the analysis on the performance of the \textit{CiaoCiao WFS} when dealing with LWE. From PASSATA simulations of MORFEO residuals, which include LWE as input together with atmospheric turbulence, LWE shows to be critical for the operation of MORFEO as it introduces differential pistons on the residual wavefront that determine an RMS of 1744 nm. Our first tests of the \textit{CiaoCiao WFS} performance when measuring the LWE-induced differential phase show promising results: the correction of the piston signals reconstructed by the \textit{CiaoCiao WFS} reduces the residual RMS of MORFEO to 256 nm.

Several aspects are currently under investigation and will be the object of future works: among these, multi-wavelength reconstruction techniques to extend the dynamic range of the WFS, integration time and spatial sampling.

\appendix    

\printbibliography 
\end{document}